\documentclass[11pt, a4paper]{article} 
\pdfoutput=1

\usepackage{jinstpub}

\usepackage{color}
\usepackage{rotating}
\usepackage{url}
\usepackage{amsmath}
\usepackage{multirow}

\title{Charge reconstruction in large-area photomultipliers}

\begin{document}

\newcommand{\pe}{\textsc{pe}}
\newcommand{\spe}{\textsc{spe}}
\newcommand{\pmt}{\textsc{pmt}}
\newcommand{\uosone}{$U_{\mathrm{OS1}}$}
\newcommand{\uostwo}{$U_{\mathrm{OS2}}$}
\newcommand{\up}{$U_{\mathrm{peak}}$}
\newcommand{\ths}{th$_{\textsc{spe}}$}
\newcommand{\thm}{th$_{\textsc{mpe}}$}
\newcommand{\specialcell}[2][c]{%
  \begin{tabular}[#1]{@{}c@{}}#2\end{tabular}}

\author[a,b,1] {M. Grassi} \note{Now at APC Laboratory - IN2P3, Paris, France}
\author[c,d,2]{M. Montuschi\note{Corresponding author}} 
\author[c,d]{M. Baldoncini}
\author[c,d]{F. Mantovani}
\author[c,d]{B. Ricci}

\author[e]{G. Andronico}
\author[a]{V. Antonelli}
\author[f,g]{M. Bellato}
\author[h,i]{E. Bernieri}
\author[a,k]{A.Brigatti}
\author[f,g]{R. Brugnera}
\author[h,i]{A. Budano}
\author[e,j]{M. Buscemi}
\author[h,i]{S. Bussino}
\author[e,j]{R. Caruso}
\author[n,o]{D. Chiesa}
\author[h,i]{D. Corti}
\author[f,g]{F. Dal Corso}
\author[a,p]{X. F. Ding}
\author[f,g]{S. Dusini}
\author[h,i]{A.Fabbri}
\author[c,d]{G.Fiorentini}
\author[q,a]{R. Ford}
\author[a,k,l,m]{A. Formozov}
\author[f]{G. Galet}
\author[f,g]{A. Garfagnini}
\author[a,k]{M. Giammarchi}
\author[f,g]{A. Giaz}
\author[e,j]{A. Insolia}
\author[f,g]{R. Isocrate}
\author[f,g]{I. Lippi}
\author[e]{F. Longhitano}
\author[e,j]{D. Lo Presti}
\author[a,k]{P. Lombardi}
\author[f,g]{F. Marini}
\author[h,i]{S. M. Mari}
\author[h,i]{C. Martellini}
\author[a,k]{E. Meroni}
\author[f,g]{M. Mezzetto}
\author[a,k]{L. Miramonti}
\author[e]{S. Monforte}
\author[n,o]{M. Nastasi}
\author[r,s]{F. Ortica}
\author[t]{A. Paoloni}
\author[a,k]{S. Parmeggiano}
\author[f,g]{D. Pedretti}
\author[r,s]{N. Pelliccia}
\author[a,k]{R. Pompilio}
\author[n,o]{E. Previtali}
\author[a,k]{G. Ranucci}
\author[a,k]{A. C. Re}
\author[r,s]{A. Romani}
\author[a,k]{P. Saggese}
\author[h,i]{G. Salamanna}
\author[f,g]{F. H. Sawy}
\author[h,i]{G. Settanta}
\author[n,o]{M. Sisti}
\author[f,g]{C. Sirignano}
\author[t]{M. Spinetti}
\author[f,g]{L. Stanco}
\author[c,u]{V. Strati}
\author[e]{G. Verde}
\author[t]{L. Votano}

\affiliation[a] {Istituto Nazionale di Fisica Nucleare, Sezione di Milano, \\Via Celoria 16, I-20133 Milano, Italy}
\affiliation[b] {Institute~of~High~Energy~Physics,\\ 19B YuquanLu, 1000049 Beijing, P.R.China}
\affiliation[c] {Dipartimento di Fisica e Scienze della Terra, Universit\`{a} di Ferrara,\\ Via Saragat 1, I-44122 Ferrara, Italy}
\affiliation[d] {Istituto Nazionale di Fisica Nucleare, Sezione di Ferrara,\\ Via Saragat 1, I-44122 Ferrara, Italy}

\affiliation[e] {Istituto Nazionale di Fisica Nucleare, Sezione di Catania,\\ Via S. Sofia 62, I-95125 Catania, Italy}

\affiliation[f] {Dipartimento di Fisica e Astronomia, Universit\`{a} di Padova, \\Via Marzolo 8 , I-35131 Padova, Italy}
\affiliation[g] {Istituto Nazionale di Fisica Nucleare, Sezione di Padova, \\Via Marzolo 8 , I-35131 Padova, Italy}

\affiliation[h] {Dipartimento di Matematica e Fisica, Universit\`{a} di Roma Tre,\\ Via della Vasca Navale 84 , I-00146 Roma, Italy}
\affiliation[i] {Istituto Nazionale di Fisica Nucleare, Sezione di Roma Tre, \\Via della Vasca Navale 84 , I-00146 Roma, Italy}

\affiliation[j] {Dipartimento di Fisica e Astronomia, Universit\`{a} di Catania,\\ Piazza Universit\'a 2, I-95131 Catania, Italy}
\affiliation[k] {Dipartimento di Fisica e Astronomia, Universit\`{a} di Milano,\\ Via Celoria 16, I-20133 Milano, Italy}
\affiliation[l] {Joint Institute for Nuclear Research, \\141980 Dubna, Russia}
\affiliation[m] {Lomonosov Moscow State University Skobeltsyn Institute of Nuclear Physics,\\119234 Moscow, Russia}

\affiliation[n] {Dipartimento di Fisica, Universit\`{a} di Milano Bicocca,\\ P.zza della Scienza 3 , I-20126 Milano, Italy}
\affiliation[o] {Istituto Nazionale di Fisica Nucleare, Sezione di Milano Bicocca, \\P.zza della Scienza 3 , I-20126 Milano, Italy}

\affiliation[p] {Gran Sasso Science Institute, \\Via Crispi 7 , I-67100 L'Aquila, Italy}
\affiliation[q] {SNOLAB,\\ Lively, ON, P3Y 1N2 Canada}

\affiliation[r] {Dipartimento di Chimica, Biologia e Biotecnologie, Universit\`{a} di Perugia, via Elce di Sotto 8 ,\\ I-06123 Perugia, Italy}
\affiliation[s] {Istituto Nazionale di Fisica Nucleare, Sezione di Perugia,\\ Via Pascoli , I-06123 Perugia, Italy}

\affiliation[t] {Istituto Nazionale di Fisica Nucleare, Laboratori Nazionali di Frascati, \\Via Fermi 40 , I-00044 Frascati (RM), 
Italy}
\affiliation[u] {Istituto Nazionale di Fisica Nucleare, Laboratori Nazionali di Legnaro, \\Viale dell'Universit\`{a} 2 , I-35020 Legnaro (PD), Italy}

\emailAdd{montuschi@fe.infn.it}

\abstract{
Large-area PhotoMultiplier Tubes (\pmt{}) allow to efficiently instrument Liquid Scintillator (LS) neutrino detectors, where
large target masses are pivotal to compensate for neutrinos' extremely elusive nature. Depending on the detector light yield,
several scintillation photons stemming from the same neutrino interaction are 
likely to hit a single \pmt{} in a few tens/hundreds of nanoseconds, resulting in several photoelectrons (\pe{}s)
to pile-up at the \pmt{} anode. In such scenario, the signal generated by each \pe{} is entangled
to the others, and an accurate \pmt{} charge reconstruction becomes challenging. 
This manuscript describes an experimental method able to address the \pmt{} charge reconstruction in the
case of large \pe{} pile-up, providing an unbiased charge estimator at the permille level up to 15 detected \pe{}s.
The method is based on a signal filtering technique (Wiener filter) which suppresses the
noise due to both \pmt{} and readout electronics, and on a Fourier-based deconvolution able to 
minimize the influence of signal distortions ---such as an overshoot.
The analysis of simulated \pmt{} waveforms shows that the slope of a linear regression modeling the relation between 
reconstructed and true charge values 
 improves from $0.769 \pm 0.001$ (without deconvolution) to $0.989 \pm 0.001$ (with deconvolution),
where unitary slope implies perfect reconstruction. A C++ implementation of the charge reconstruction algorithm is available 
online at~\cite{website}.
}


\maketitle
\flushbottom


\section{Introduction}

The consolidated Liquid Scintillator (LS) technology is driving neutrino physics into the era of precision calorimetry. The unprecedented scientific achievements of Borexino~\cite{borexino}, Daya Bay~\cite{dayabay}, Double Chooz~\cite{doublechooz}, KamLAND~\cite{kamland} and RENO~\cite{reno} experiments have been the trailblazer for a new generation of multi-kiloton detectors (JUNO~\cite{juno}, Jinping~\cite{jinping}, RENO50~\cite{reno50}, SNO+~\cite{snoplus}, ANDES~\cite{andes}). Going beyond the open issues in particle physics, the perspective will be some spin-off in applied antineutrino physics \cite{bowden}.

The experimental challenges of neutrino calorimetry revolve around improving both the energy and the spatial resolution of (anti)neutrino interaction detection. 
Since both resolution terms scale with the number of detected scintillation photons, maximizing the detector photocoverage ---namely the sensitive area of the detector--- is pivotal to improve them. In many current and future detectors, technical and budget constraints make the use of large-area  PhotoMultiplier Tubes (\pmt{}) the only viable solution to achieve large photocoverage. In some cases, these constrains even justify a R\&D program dedicated to develop a novel \pmt{} technology~\cite{mcppmt}.  The largest \pmt{} bulbs built to date are 20-inch diameter. Because of their large acceptance,
they typically detect many photoelectrons (\pe{}) per scintillation event, which are likely to pile up at the readout level. That is, the spacing between the \pe{}s' time of arrival is lower or comparable to the width of a single-photoelectrons (\spe{}) pulse. 

When \pe{} pulses overlap, their identification becomes challenging. 
Especially if the pulses are affected by an overshoot (a distortion in the \pmt{} output signal described in section~\ref{sec2}), two subsequent \pe{}s could easily be mis-reconstructed as a single one.
A biased \pmt{} charge reconstruction not only compromises the linearity ---and therefore the resolution--- of the detector-wise energy estimator, but also threatens the time-based reconstruction of the event vertex. In a large detector, a precise knowledge of the event vertex is crucial to define a fiducial volume meant to reject background energy depositions arising from natural radioactivity. 
The relevance of this issue can be further appreciated by noting that the Daya Bay experiment, after 4 years of smooth data taking, 
developed a new readout system based on fast digitizers, and its associated waveform reconstruction, to better assess the linearity of the detector energy response~\cite{dayabay_fadc}. 

The goal of this study is twofold: (\textsc{i}) to propose an open-source detector-independent charge reconstruction algorithm~\cite{website} processing the output of a generic fast digitizer (FADC) connected to a \pmt{}, and (\textsc{ii}) to define a procedure to assess the accuracy of the reconstructed charge, especially in the case of large \pe{} pile-up. While (\textsc{i}) is based on realistic signal and noise assumptions, and particular care was devoted to model most of the \pmt{} peculiarities, (\textsc{ii}) is meant to allow a comparison between any charge reconstruction algorithm. It is worth mentioning that our algorithm is specifically designed to minimize those charge reconstruction biases introduced by the presence of an overshoot in the \pe{} pulses, and by the noise fluctuations embedded in the  \pmt{} output pulse.


\section{PMT  Waveform Simulation}
\label{sec2}


\begin{figure}[t]
\center
\includegraphics[width=\textwidth]{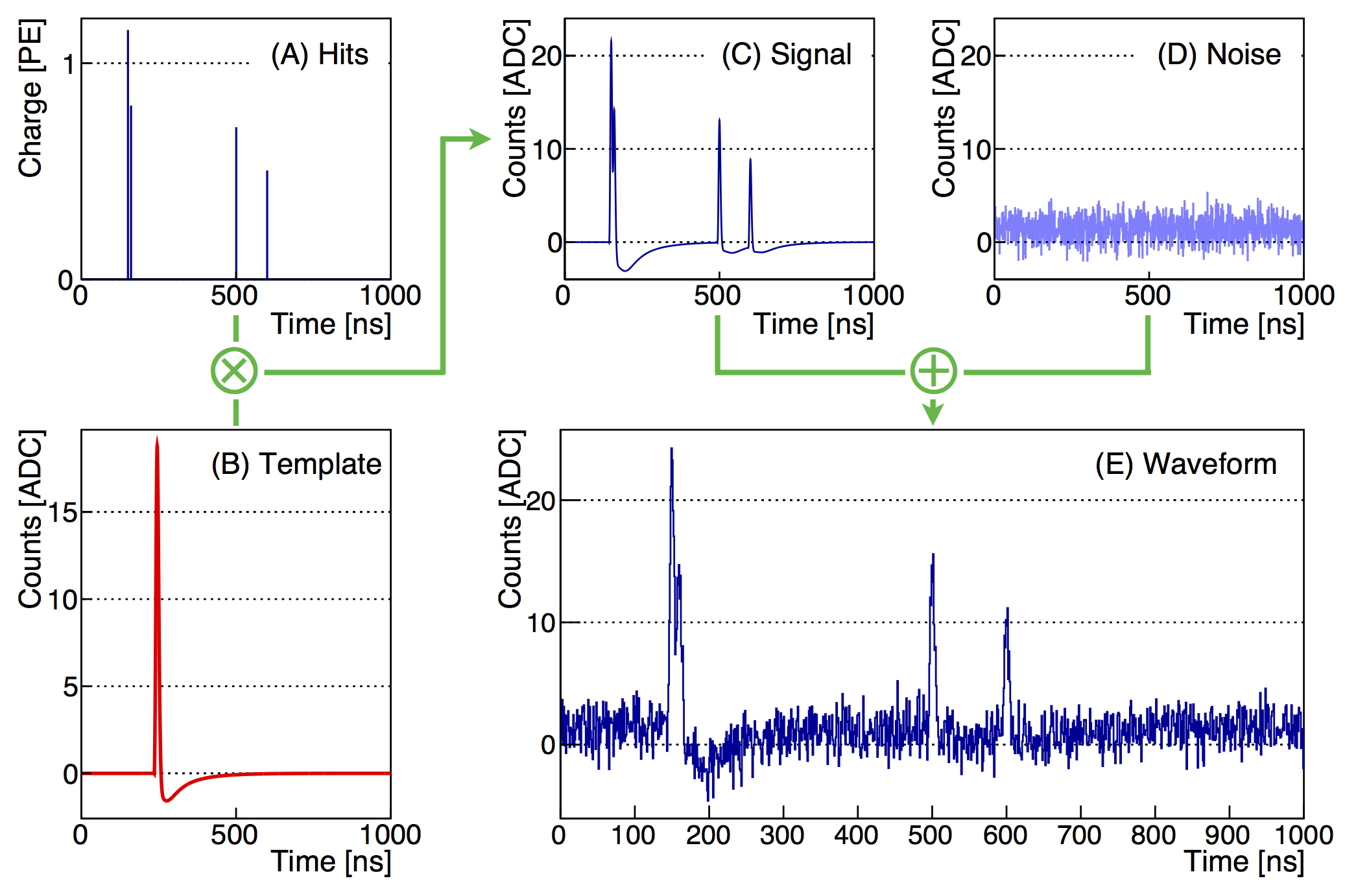}
\caption{
Building blocks of the waveform simulation. Panel A shows the hit generation, where for each detected \pe{} 
a random charge value is generated. Panel B shows the analytical shape of a single \pe{} template. Panel C is
the result of the convolution between the \spe{} template and the hits. Panel D contains the \pmt{} simulated noise.
Panel E shows the final \pmt{} waveform resulting from the sum of the signal and of
the noise components.}
\label{fig1}
\end{figure}

We build a \pmt{} waveform simulation with the aim to develop and validate the charge reconstruction algorithm with a known input signal.
To avoid to rely on the specification of a given manufacturer and/or model, we implement a parametric simulation of a generic \pmt{} response. The emission of a \pe{} from the photocathode, its collection by the anode, and its subsequent amplification are simulated as an instantaneous \pmt{} charge output.
In accordance to experimental data~\cite{jetter}, the charge value $q$  is generated randomly according to the distribution $f(q)$ in eq.~\ref{eq:spe_distro}, adapted from~\cite{dossi}
\begin{equation}
\label{eq:spe_distro}
f(q) =  
\begin{cases} 
(1-\omega) \frac{1}{\sigma \sqrt{2 \pi}} \exp \Big[-\frac{(q-q_0)^2}{2 \, \sigma^2}\Big] + \frac{\omega}{\tau} \exp{- \frac{q}{\tau}} & q \geq q_p \\
0 & q < q_p
\end{cases}
\end{equation}
The weight $\omega$ determines the relative contribution of a Gaussian distribution with respect to an exponential tail, modeling the faction of under-amplified \pe{}s. The width of the former ($\sigma$) is set to 30\% to describe the typical uncertainty induced by the first dynode amplification ($\sim 1/\sqrt{10}$). The Gaussian mean value ($q_0$) is set to 1, as in the case of a perfect gain calibration. The exponential tail accounts for low amplitude hits due to elastically scattered and backscattered electrons from the first dynode \cite{dossi}. The parameter $q_p$ sets the threshold for the minimum charge to be considered in order to avoid to simulate the pedestal. The values of all the parameters used in eq.~\ref{eq:spe_distro} are listed in table~\ref{tab1}, and the analysis of their variability is described in section~\ref{sec4}. The time at which a charge output occurs is randomly sampled from a flat distribution spanning a $1~\mu$s-long time window, and each charge-time pair is defined as a \textit{hit}.


\begin{table}[t]
\caption{Parameters used in eq.~\ref{eq:spe_distro} together with their nominal values.}

\center
\begin{tabular}{ |c   c  c  |}
\hline
Parameter  & Description & Value \\
\hline
q$_o$  & \spe{} calibrated gain&  1 \pe{} \\
q$_p$ & Pedestal cutoff & 0.3 \pe{}  \\
$\sigma$ & \spe{} Gauss width & 0.3 \pe{}  \\
$\omega$ & Under-amplified \pe{} fraction & 0.2  \\
$\tau$  & Exponential decay constant  & 0.5 \pe{}    \\  
\hline
\end{tabular}
\label{tab1}
\end{table}


\begin{table}[t]
\caption{ 
Parameters involved in the signal (top rows) and noise (bottom rows) simulation of the \pmt{} waveform. 
Parameters for which the ``Range'' column is filled are generated randomly according to a flat probability density function at the beginning of the simulation.
}
\center
\begin{tabular}{|ccccc|}
\hline
Parameter & Value & Range & Description & Function \\
\hline
 $U_0$       & 20 ADC  & & \multirow{3}{*}{\spe{} template} & \multirow{3}{*}{eq.~\ref{eq3}} \\ 
 $\sigma_0$  & 0.15     & & &                      \\ 
 $\tau_0$    & 30 ns  & & &                      \\
\hline
 $U_1$       & -1.2 ADC & & \multirow{3}{*}{Overshoot} & \multirow{3}{*}{eq.~\ref{eq4}} \\ 
 $\sigma_1$  & 55 ns        & & &                      \\ 
 $t_1$       & -4 ns   & & &                      \\ 
\hline
 $U_2$       & -2.8 ADC  & & \multirow{2}{*}{Overshoot} & \multirow{2}{*}{eq.~\ref{eq5}} \\
 $\tau_2$    & 80 ns     & & &  \\

\hline
\hline

 $\mu_{\mathrm{N}}$    & 1.5 ADC  &      & Baseline & \multirow{2}{*}{Gaussian}            \\
 $\sigma_{\mathrm{N}}$ & 1 ADC &      & White Noise&                                   \\
\hline
 $n_{\mathrm{FF}}$     & 5 &                & N(Components) & \multirow{3}{*}{\specialcell{Fixed\\Frequency}}  \\
 $f_{\mathrm{FF}}$     &    & [1, 480] MHz   & Frequency                           & \\
 $A_{\mathrm{FF}}$     &    & [0.1, 0.3] ADC & Amplitude                           & \\

\hline
\end{tabular}
\label{tab2}
\end{table}

A sequence of hits, resulting from several \pe{}s impinging on the same \pmt{}, are shown  as instantaneous pulses in figure ~\ref{fig1}A. 
The pulse position on the horizontal axis represents the hit time with respect to a reference $t_0$ ---for example a global detector trigger---, and 
the pulse amplitude represents the \pmt{} output charge.
We assume the \pmt{} to be connected to a fast electronics able to sample its output at 1~GSample/s, and we describe the voltage drop
resulting from the detection of each hit
with a log-normal function (eq.~\ref{eq3}) \cite{jetter,chess}. The AC coupling often used to split the \pmt{} high-voltage from the \pmt{} output signal might induce a distortion in the \pmt{} output waveform. We refer to the component of the signal with a polarity opposite to the log-normal as the overshoot, and we follow~\cite{jetter} to describe it using the sum of a Gaussian (eq.~\ref{eq4}) and an exponential tail (eq.~\ref{eq5}). The values of the parameters used in eq.~\ref{eq3},~\ref{eq4} and~\ref{eq5} are listed in table~\ref{tab2}. The values related to the \spe{} amplitude are chosen to obtain an approximate 0.5 mV/count resolution, resulting in a dynamic range of $\sim$50\pe{} when using a commercial 10-bit digitizer. The overshoot values are adapted from the Daya Bay experience~\cite{jetter}.  
The analytical shape of a \spe{}-waveform 
is reported in eq.~\ref{eq:spe} and shown in figure ~\ref{fig1}B. We name this shape the  \spe{} template. The signal-only waveform resulting from the hits shown is  figure ~\ref{fig1}A is built by convolving each hit with the  \spe{} template, and it is shown in figure ~\ref{fig1}C.

\begin{equation}
U_{\mathrm{peak}}(t)=U_0 \cdot  \exp \left(-\frac{1}{2} \left(\frac{\ln(t/\tau_0)}{\sigma_0}\right)^2 \right)
\label{eq3}
\end{equation}
\begin{equation}
U_{\mathrm{OS1}}(t)=U_1 \cdot  \exp \left(-\frac{1}{2} \left(\frac{t-t_1}{\sigma_1}\right)^2 \right)
\label{eq4}
\end{equation}
\begin{equation}
U_{\mathrm{OS2}}(t)=U_2 \cdot \frac{1}{\exp\left(\frac{50\,\mathrm{ns}-t}{10\,\mathrm{ns}} \right) +1 } \cdot \exp \left(-\frac{t}{\tau_2}  \right)
\label{eq5}
\end{equation}
\begin{equation}
\label{eq:spe}
U(t) = U_{\text{peak}} + U_{\text{OS1}} + U_{\text{OS2}}
\end{equation}

To make the simulation more realistic, we add a noise waveform (figure ~\ref{fig1}D) to the waveform built using only signal hits. The simulated noise includes a Gaussian component and some periodic components with fixed frequency. The former describes  an overall baseline offset and time-uncorrelated baseline fluctuations. The Gaussian mean ($\mu_{\mathrm{N}}$) is set to 1.5~ADC counts, and the Gaussian width ($\sigma_{\mathrm{N}}$) to 1~ADC count. The fixed-frequency components describe potential noise sources embedded in, or due to, the readout circuit. We simulate 5 such components ($n_{\mathrm{FF}}$), with a random amplitude generated flat in the 0.1-0.3 ADC counts range, and with a random frequency in the 1-480~MHz range. The ultimate waveform comprising signal hits and all noise components is shown in  figure ~\ref{fig1}E.


\section{Charge Reconstruction}
\label{sec3}


\begin{figure} [h]
\center
\includegraphics[width=\textwidth]{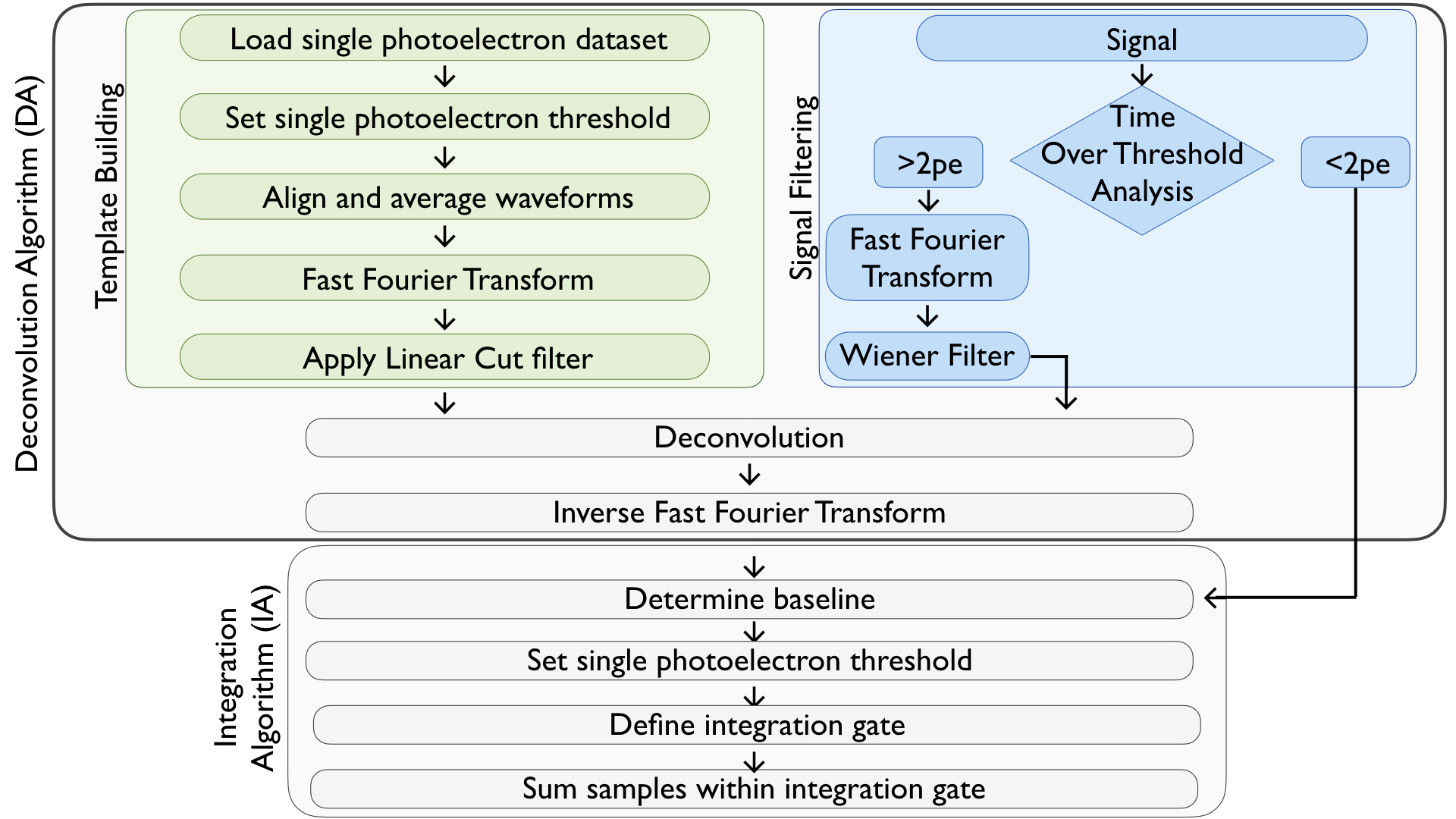}
\caption{
Schematic diagram of the charge reconstruction algorithm. 
The \spe{} template in the time domain is computed by averaging 10$^4$ \spe{} 
waveforms that simulate a LED-based calibration dataset. 
A Fast Fourier Transform (FFT) is applied to both the \spe{} template and the \pmt{} 
waveform  to be reconstructed.
The latter is processed using the Wiener Filter to minimize the noise and suppress the 
overshoot. 
The \spe{} template in the frequency domain is then used as a benchmark pattern 
to deconvolve the filtered waveform in the frequency domain. 
We eventually process the deconvolved waveform with an Inverse FFT,
so that the waveform in the time domain could be integrated to compute
the \pmt{} output charge.
}
\label{fig2}
\end{figure}


\begin{figure}[t]
\centering
\includegraphics[width=0.7\textwidth]{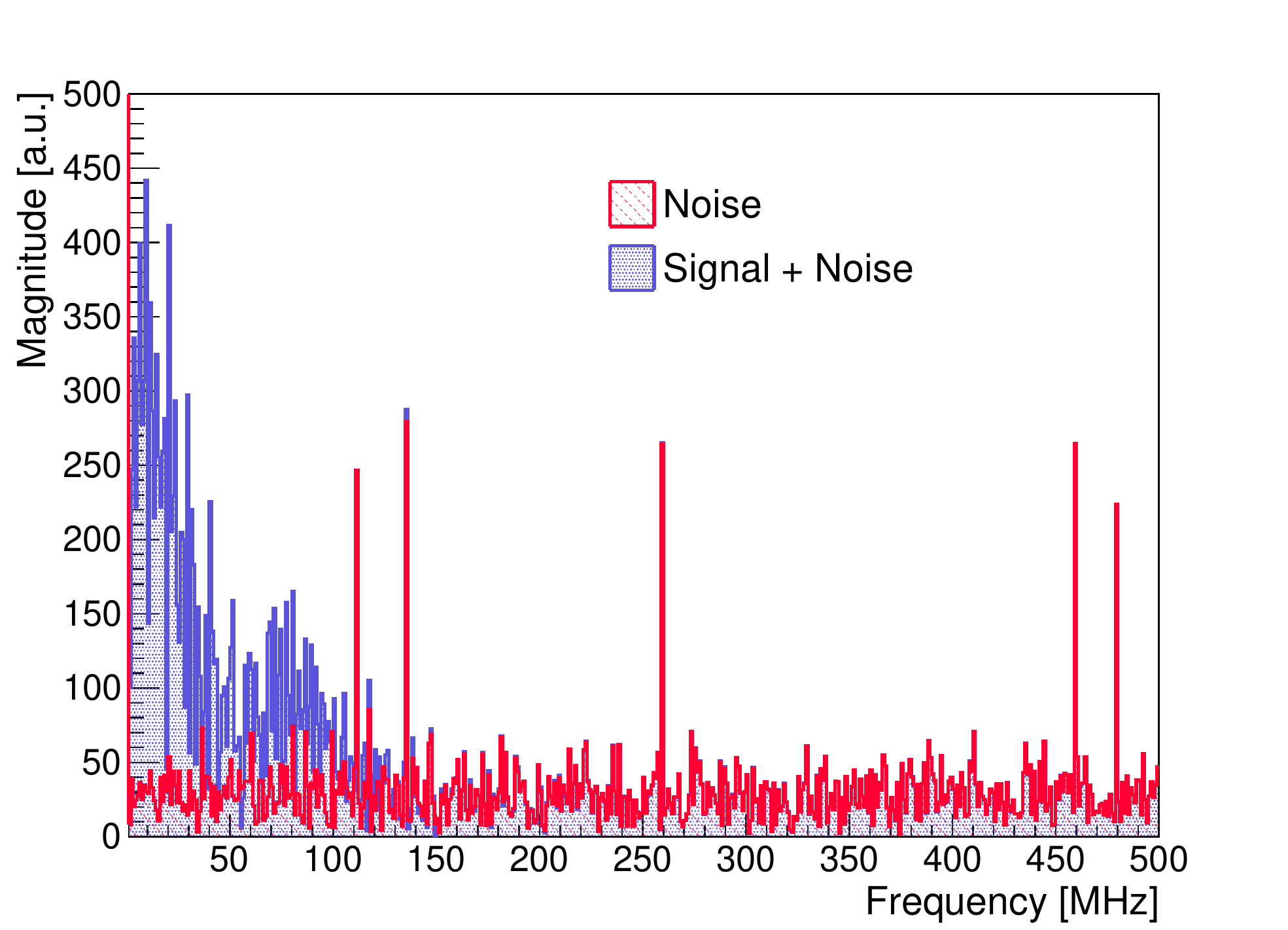}
\caption{
Frequency power spectra of a pure-noise waveform (red) and a complete waveform comprising noise and \pe{} pulses (blue). The two spectra are used in eq.~\ref{eq13} to derive the Wiener Filter. The spikes in the noise spectrum are due to the fixed-frequency noise components. 
}
\label{fig3}
\end{figure}


\begin{figure}[t]
\center
\includegraphics[width=0.7\textwidth]{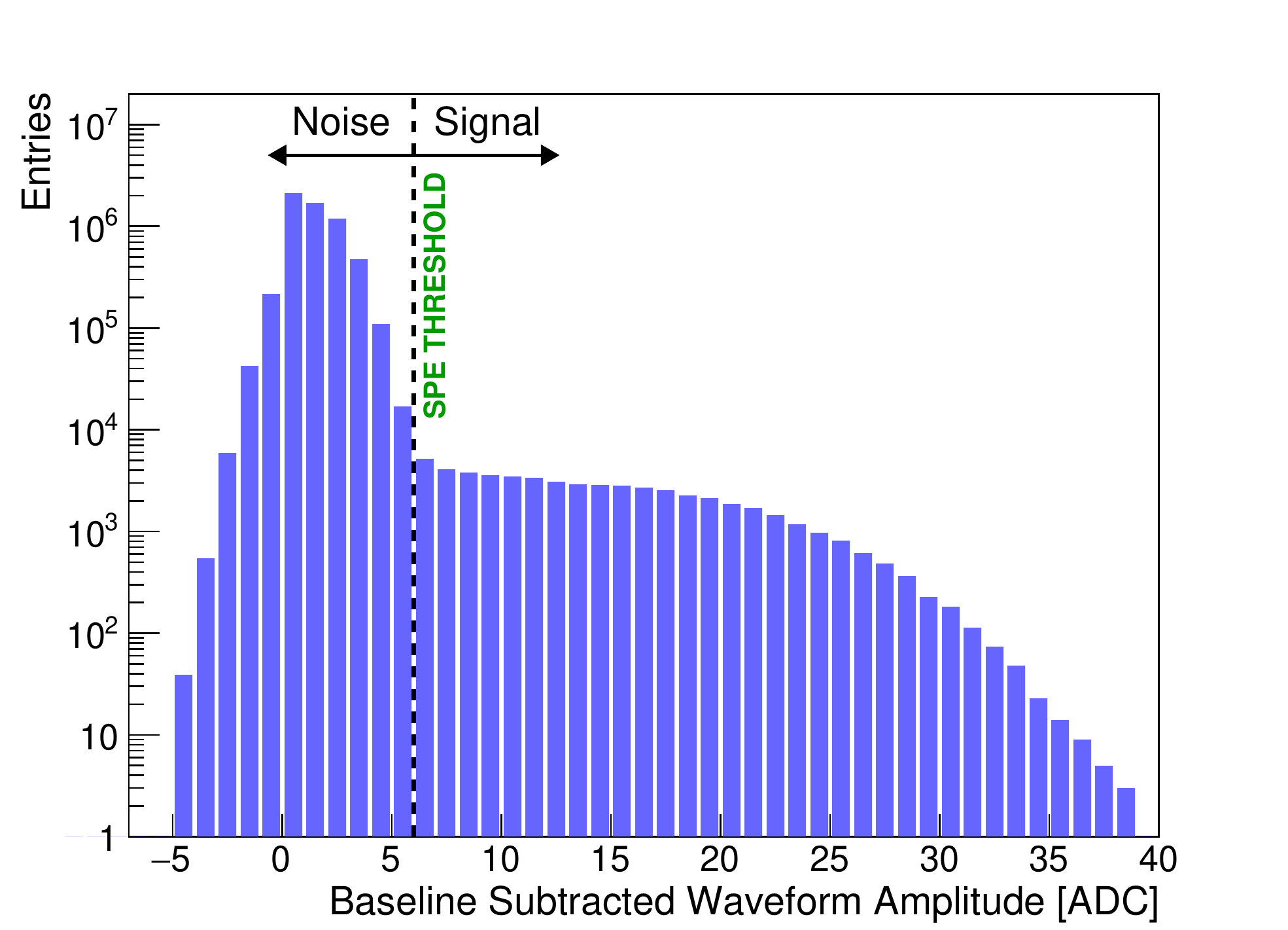}
\caption{
Distribution of all the baseline-subtracted FADC samples in a waveform, built using $10^4$ waveforms containing mostly 0 or 1 hit (dataset S). 
Noise fluctuations are responsible for the peak centered at zero. FADC samples related to the fraction of the waveform where \spe{} pulses occurred
are responsible for the tail extending to large ADC values. A 6 ADC trigger threshold is chosen to identify waveforms where at least 1 hit is present.}
\label{fig4}
\end{figure}

The simplest way to determine the \pmt{} output charge is to sum up all the waveform samples collected in a readout window. However, this approach would embed both the noise and the overshoot into the charge measurement, resulting in a rough and biased charge estimator. The charge reconstruction algorithm described in this section and sketched in figure~\ref{fig2} is meant to mitigate the role of both noise and overshoot.
It comprises two steps: (\textsc{i}) 
the Deconvolution Algorithm (DA), which filters
 the raw waveforms reducing the noise, and
deconvolves \spe{} templates out of the filtered waveforms; and
(\textsc{ii}) the Integration Algorithm (IA), which integrates the deconvolved waveforms
with the aim to determine the overall \pmt{} output charge.
 
Both filtering and deconvolution are data-driven methods that need to be trained and tested. To this end, we rely on waveforms generated using the simulation package described in section~\ref{sec2}. We produce two large datasets, one comprising mostly \spe{} waveforms (S dataset), and one with waveforms containing multiple hits (M dataset). In order to emulate a real experimental setup, the charge reconstruction algorithm is designed not to use the true hit information (time and charge). We unblind such parameters only when assessing its performance ---as described in section~\ref{sec4}. The first dataset emulates the behavior of a low-intensity LED placed in front of a \pmt{} to measure its \spe{} response, where one typically gets either 0 hits or a single hit in the resulting waveform (p(0)$\simeq$0.3, p(1)$\simeq$0.6, and p($>$1)$\simeq$0.1). In the second dataset, waveforms are generated such that the number of true hits follows a flat distribution in the [0-15] range. This distribution is chosen to investigate the performance of the reconstruction algorithm even in those events experiencing high pile-up.

The noise pattern present in simulated waveforms consists mostly of bin-to-bin uncorrelated fluctuations. On the contrary, a \spe{} pulse lasts for a few tens of ns. The difference between the two time scales is highlighted in figure~\ref{fig3}, where the frequency spectra of a waveform with no hits (pure noise, red curve) and a waveform with 4 hits (blue curve) are compared. The former is rather flat, with the exception of the fixed-frequency noise components, which manifest as spikes in the frequency domain. The latter, on the contrary, is peaked at low frequencies, as a result of \spe{} pulses being much slower than the noise fluctuations. We exploit such a difference to suppress the noise by means of a Wiener Filter ---a technique commonly employed in signal processing~\cite{steven}. 
The filter is defined by the following kernel equation 	
\begin{equation}
H[f]= \frac{|S[f]|^2}  {  |S[f]|^2 + |N[f]|^2 }
\label{eq13}
\end{equation}
where $S[f]$ and $N[f]$ are the frequency spectra of the signal and of the noise.
In literature, this filter is referred to as the \textit{optimal} linear filter for the removal of additive noise~\cite{wiener_description_6}. Namely, 
the coefficients $H[f]$ of the Wiener Filter are calculated to minimize the average squared distance between the filter output and the desired 
signal~\cite{wiener_mit}.
Such coefficients are then used to weight the frequency components of the waveform being processed.

Eq.~\ref{eq13} can be rearranged by dividing both the numerator and the denominator by the noise power spectrum $|N[f]|^2$, and by substituting the variable 
$\text{SNR}[f] = |S[f]|^2 / |N[f]|^2$. This manipulation yields 
\begin{equation}
H[f] = \frac{\text{SNR}[f]}{\text{SNR}[f] + 1}
\label{eq:wiener_snr}
\end{equation}
where SNR  is a ratio of the signal power to the noise power.
Eq.~\ref{eq:wiener_snr} 
makes the frequency response of the filter more intuitive, being a real positive number in the range $0 \leq H[f] \leq 1$.
Frequencies that are barely affected by noise ($\text{SNR}[f] \to \infty$) result in the filter being close to unity, hence
applying little or no attenuation to the input components. On the contrary, frequencies that are severely affected by noise 
($\text{SNR}\sim 0$)  result in the filter to heavily attenuate them ($H[f]\sim 0$). In summary, the Wiener Filter attenuates each frequency 
component in proportion to an estimate of its signal-to-noise ratio.

The deconvolution is a technique meant to identify and resolve the presence of \spe{} pulses within a filtered waveform, with the aim to perform an unbiased  measurement of the \pmt{} output charge. It is implemented as a division in the frequency domain between the waveform to be reconstructed and the template of a \spe{} pulse. 
We build the template by time-aligning and averaging $10^4$ \spe{} waveforms selected from the S dataset, which emulates calibration data collected by illuminating the \pmt{} with a low intensity LED.
The selection determining which waveforms are to be used in the template building is based on a threshold-crossing criterion: only waveforms whose baseline-subtracted amplitude exceeds 30\% of the mean \spe{} amplitude are retained. The distribution of  all the digitized samples in the S dataset (1000 samples per waveform) is shown in  figure~\ref{fig4}, where the selection threshold clearly marks the transition between the noise region and the signal region. 
To further clean the  \spe{} template from those noise contributions not suppressed by the averaging process, all frequencies above 120~MHz are stripped. 

\begin{figure}[t]
\center
\includegraphics[width=0.8\textwidth]{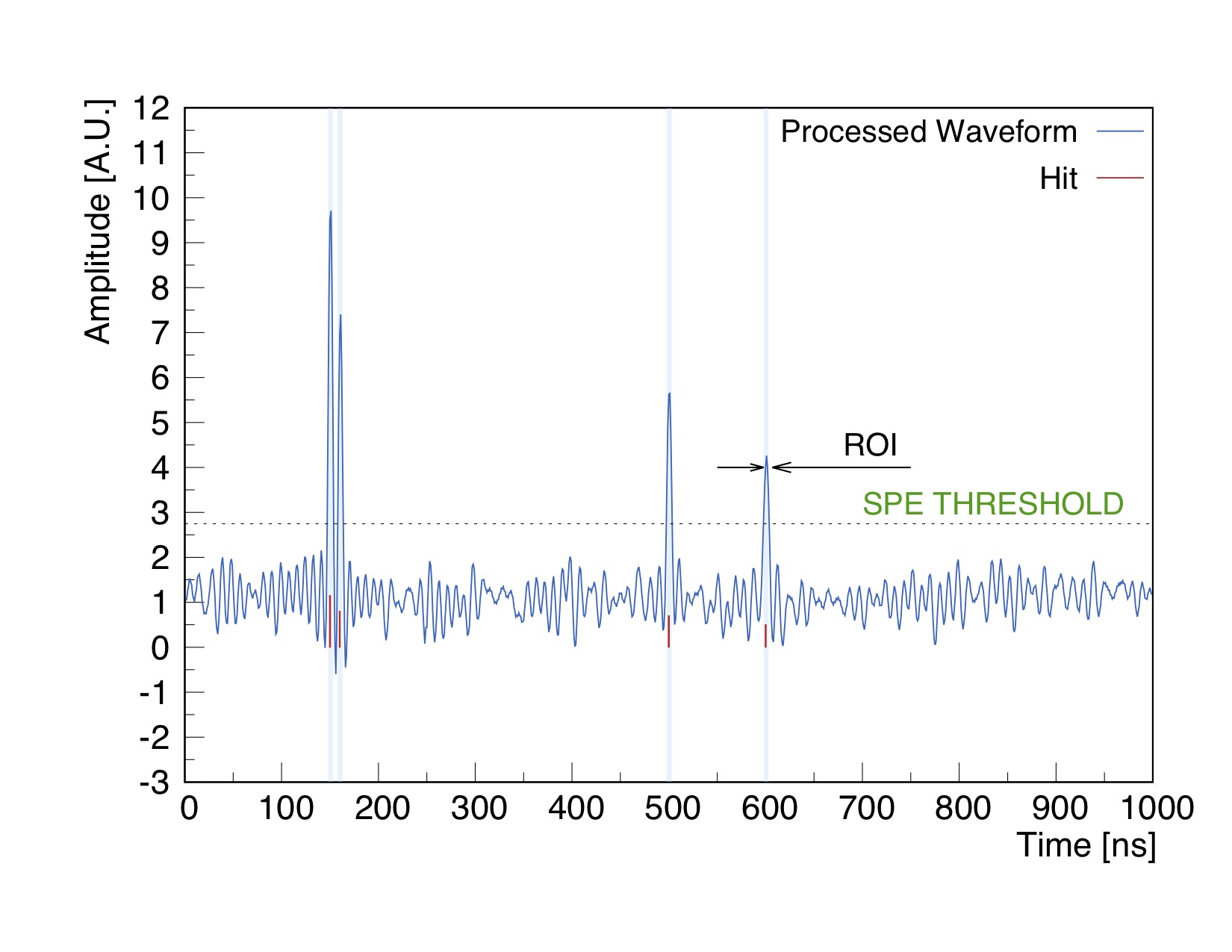}
\caption{Waveform arising from filtering and deconvolving figure~\ref{fig1}E. True hits are shown for reference in red. The charge of each \pe{} is reconstructed by integrating the bins falling within a ROI (shaded areas) defined starting from the time at which the waveform crosses a trigger threshold (dashed line). The threshold is set to be 30\% of the amplitude of a \spe{} pulse.  }
\label{fig5}
\end{figure}

The outcome of the deconvolution performed on the waveform in figure~\ref{fig1}E is shown in figure~\ref{fig5}. Here the deconvolved waveform is brought back to the time domain by means of an Inverse Fast Fourier transform, and it shows narrow pulses whose amplitude is proportional to the original hit charge.   However, rather than using the pulse amplitude, we find a better estimator of the hit charge to be the integral of the pulse. 

To minimize the contribution of residual noise to the pulse integral, and the possible biases arising from the ringing visible in the vicinity of the pulses (Gibbs effect), only waveform samples falling within a Region Of Interest (ROI) are summed up to yield the final charge value. A ROI is opened any time the deconvolved waveform crosses a threshold 
corresponding to 30\% of the amplitude of a deconvolved \spe{} pulse. 
We define the minimal ROI to be 6~ns wide, and we extend it if at its end the waveform is still 
above the threshold. 

Following an extensive optimization procedure, we determine that our reconstruction algorithm is better suited to process waveforms with more than two hits. Below this value, the CPU power required to filter and deconvolve the waveform is not paying off in terms of charge reconstruction performance. As a consequence, we implement a waveform preselection to determine how to process each waveform. We define $\Delta t$ to be the time during which the \spe{} template stays over threshold (6~ns). When a waveform is found to be over threshold for at least $2\,\Delta t$, it undergoes the full charge reconstruction comprising filtering and deconvolution. Otherwise, a simple integral of the bins over threshold is used to compute the reconstructed charge.


\section{Results and discussion}
\label{sec4}

The performance of the charge reconstruction algorithm 
is assessed by processing a set of $10^4$ simulated waveforms 
containing a random number of \pe{} in the range [0,15], namely the M  dataset. 
In particular we aim to show how the algorithm improves the precision and accuracy of the charge 
estimate with respect to a simple integration.

\begin{figure}[t]
\centering
\includegraphics[width=0.9\textwidth]{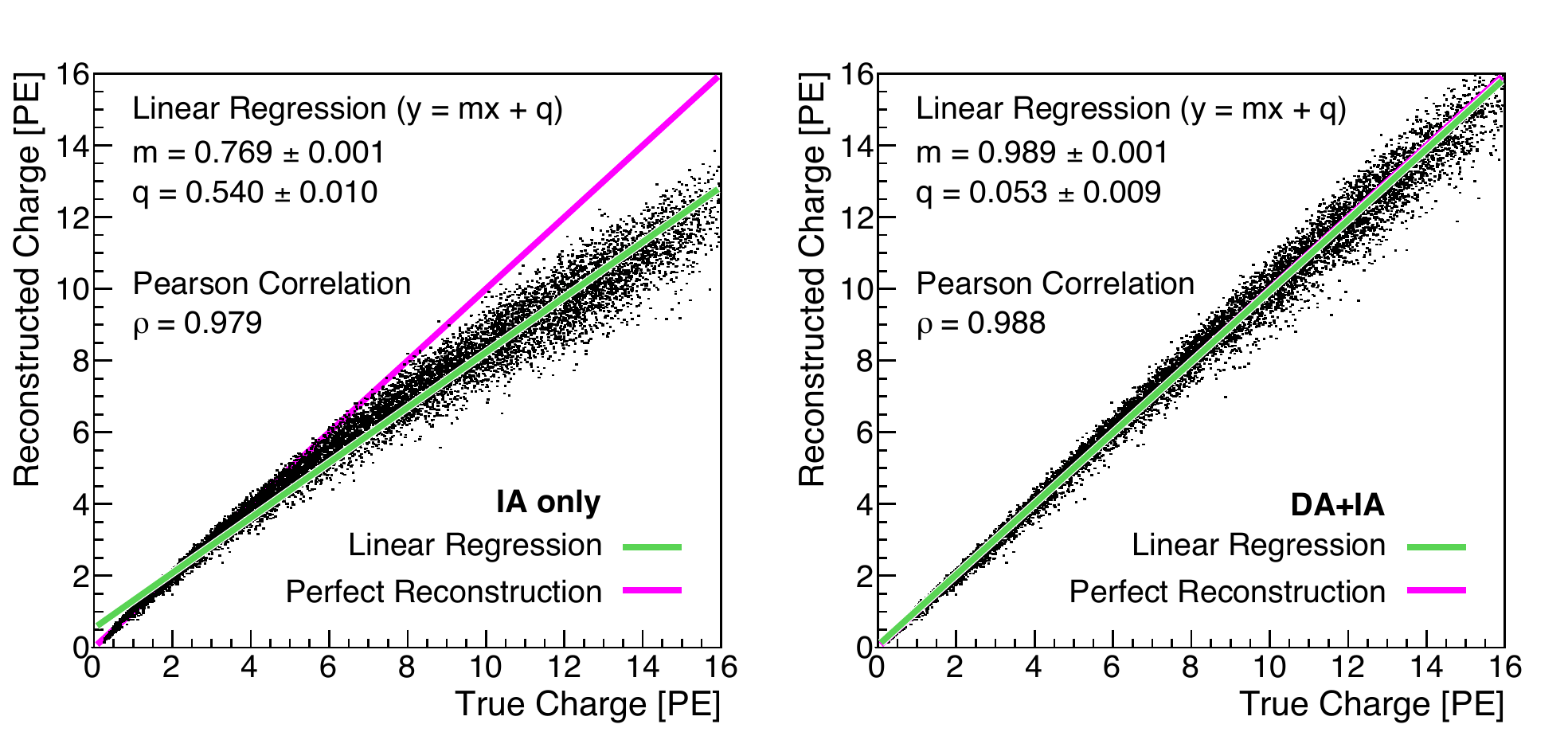}

\caption{
Reconstructed charge versus true charge for $10^4$ \pmt{} waveforms.  The reconstructed charge is computed using the IA alone (left) and DA+IA (right). In both panels the green line is the result of a linear regression performed on the $10^4$ data points.}
\label{fig6}
\end{figure}

\begin{figure}[htb]
\center
\includegraphics[width=\textwidth]{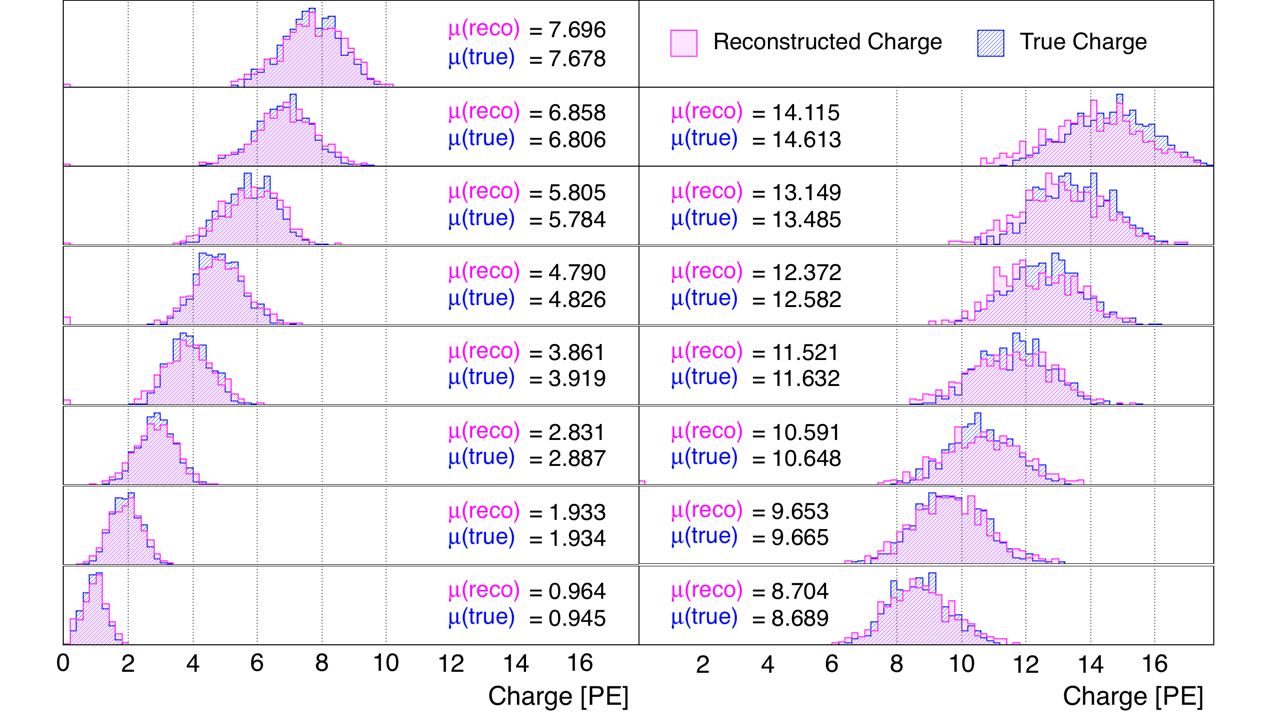}
\caption{Distribution of the true charge (blue) and of the reconstructed charge (pink) using the full reconstruction algorithm (DA + IA). Each plot is built using waveforms with a defined number of \pe{}s ranging from 1 (bottom left) to 15 (top right). The mean true charge is consistently lower than the true number of \pe{}s because of the exponential tail used to simulate the \spe{}
distribution in eq.~\ref{eq:spe_distro}.}
\label{fig:histo}
\end{figure}

To quantify the performance of the algorithm, we process 
the M dataset waveforms twice, once using the IA alone, and once using 
both the DA and the IA. For each of the two reconstruction approaches we build
a correlation plot using pairs of reconstructed and true charge values,
as shown in  figure~\ref{fig6}.
An ideal reconstruction algorithm would result in the two quantities to 
be maximally correlated, and a linear regression would yield unitary slope
and null intercept.
Any deviation from such behavior is therefore to be interpreted as a bias in the reconstruction.
In particular, a non-null intercept ($q$) models any bias that does not depend on the number of
reconstructed \pe{}, while a non-unitary slope ($m$)
models any \pe{}-dependent bias effectively compromising the linearity of the reconstruction 
algorithm. 
To further measure how scattered the reconstructed charge values are 
with respect to the true charge values, we use the Pearson correlation coefficient ($\rho$).
We interpret the latter as an estimator of the charge reconstruction precision,
while we interpret the linear regression coefficients as a measurement of its accuracy.
By comparing the two plots shown in  figure~\ref{fig6}, it can be noticed that the algorithm 
significantly improves both the precision and the accuracy of the charge reconstruction.
Indeed, $m$ improves from 0.769 to 0.989, $q$ improves from 0.540
to 0.053, and $\rho$ improves from 0.979 to 0.988.
To make the meaning of these numbers more evident, we report that 
if waveforms with 5 and 10 true \pe{}s are processed with the IA alone,
the average reconstructed charge is biased by 7\% and 12\% respectively.
While, in the case of DA+IA, the bias becomes negligible (at permille level)
in both cases. Figure~\ref{fig:histo} additionally shows the distribution of the true and reconstructed charge values 
for events with a number of \pe{}s ranging between 1 and 15.

\begin{figure}
\center
\includegraphics[width=0.8\textwidth]{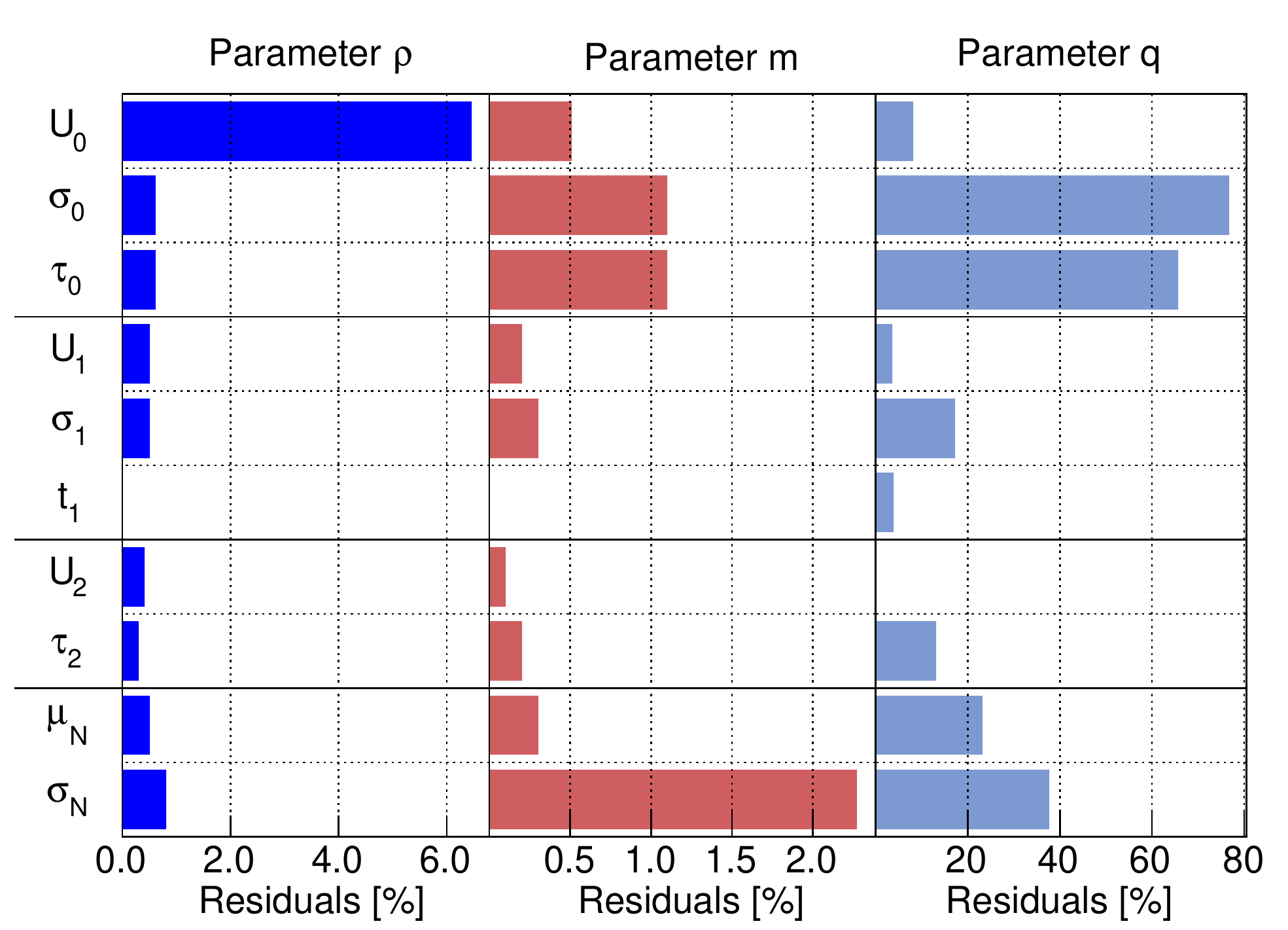}
\caption{Residuals of the three performance parameters (linear regression coefficients and correlation coefficient) assessing how the charge reconstruction worsens when the input waveforms contain pulses that are different from the \spe{} template. The input waveforms here are produced by applying a 10\% smearing to each of the parameters listed on the vertical axis of the plot.  }
\label{fig_sys}
\end{figure}

We further assess the resilience of the reconstruction algorithm to possible distortions in the input waveform. 
The data taking of a typical neutrino experiment spans indeed several years ---if not decades---, during 
which several aging issues might compromise the initial \pmt{} performance. Such issues often affect 
both the noise level and the shape of \spe{} pulses. 
To evaluate how  the precision and the accuracy of the charge reconstruction degrades due
to variations in the \pmt{} waveforms with varying shape, we produce new datasets in which all the
\pmt{} input parameters are smeared by 10\%, one at a time.
In analogy to the procedure described above, (\textsc{i}) we process all these datasets
using the full reconstruction algorithm (IA+DA); (\textsc{ii}) for each dataset we build the 
reco-true charge correlation plots; (\textsc{iii}) we use these plots to perform a linear regression and to compute the 
Pearson correlation coefficient. The discrepancy between the resulting values and the nominal values
are plotted in terms of residuals in figure~\ref{fig_sys}. Some considerations follow.

\begin{itemize}
\item Given the different role that $\rho$, $m$ and $q$ play in assessing the algorithm performance, only residuals 
within the same panel can be compared. That is, residuals here are meant to draw a hierarchy among the 
parameters describing the \spe{} shape, with the aim to show which of them affects the reconstruction algorithm the most. 

\item From the left panel it is evident that injecting \spe{}s with varying amplitude ($U_0$) compromises
the algorithm precision. This is expected, since the \spe{} template now fails to describe the \spe{} pulses 
present in the \pmt{} output waveforms.

\item The parameters defining the width of the 
\spe{} shape, namely $\tau_0$  and $\sigma_0$, heavily affect the accuracy of the linear regression parameters.

\item The parameters shaping the overshoot ($U_1$, $\sigma_1$, $t_1$, $U_2$, $\tau_2$) 
play a negligible role, as a consequence
of the successful overshoot stripping by the DA. 

\item The white noise amplitude
($\sigma_N$) is able to introduce a non-negligible charge bias, suggesting that the implementation of
the filter has room to be improved. 

\end{itemize}


\section{Conclusions}
\label{sec5}

This manuscript describes a new method to reconstruct the output charge of a \pmt{} when sampled with a fast digitizer.
Its  originality stands in putting together several well established signal processing techniques with the aim to improve the charge reconstruction accuracy over a large dynamic range, in terms of both mean value of the reconstructed charge and of dispersion around the mean. 

The algorithm comprises a filtering step, a deconvolution step, and an integration step, which aim to reduce the \pmt{} noise, 
to compensate for predictable distortions in the \pmt{} waveform, and to infer the number of \pe{}s detected by the \pmt{}.
Some of its features are summarized here below.

\begin{itemize}

\item The filter allows to mitigate not only white noise, but also noise at fixed frequencies
 often introduced at the level of the readout electronics (e.g. due to grounding issues).

\item The deconvolution is based on a \spe{} template, which can be automatically derived from dark count 
events, effectively introducing a zero dead-time calibration procedure. In the case of an experiment 
using different \pmt{} types, the algorithm can handle all of them naturally by computing a 
different template for each type. Moreover, the capability to build new templates continuously 
allows to account for any variation in the \pmt{} performance over time.

\item The deconvolution allows an a-priori estimate of the time needed to reconstruct each waveform. 
The time needed to perform the most CPU-intensive operation (Fast Fourier Transform) 
depends only on the number of samples within a waveform, and not on its complexity.
On the contrary, an approach based on reconstructing the charge by means of an analytic fit 
would become slow and unreliable in case of large pile-up. 
Such consideration becomes even more relevant in detectors instrumented 
with a large number of \pmt{}s.

\item The deconvolution effectively reduces the undesired \spe{} features from the \pmt{}
waveform (such as the overshoot), which can in principle bias the charge reconstruction, 
making the integration step much more robust.

\end{itemize}

We tested the reconstruction algorithm by analyzing waveform datasets produced with a custom-made \pmt{} simulation.
The latter allowed us to simulate a generic \pmt{} readout starting from an analytic \spe{} signal template.

We investigated the effect of processing waveforms with a shape different from the \spe{} template. We did it by producing
a sample of waveforms where the parameters describing  the \spe{} pulses were smeared randomly by 10\% around their nominal value. The aim was to test the resilience of the algorithm to any change in the \pmt{} waveform due, for instance, to aging issues.
We determined that the reconstruction algorithm is differently sensitive to the input parameters.
In particular, those describing the amplitude and the duration of a \spe{} pulse are the most likely to bias the reconstruction performance.

The overall charge reconstruction algorithm here described is geared to play a key role 
in improving the energy resolution of LS detectors with large photocoverage.
Such detectors are indeed expected to be severely affected by energy-related systematic uncertainties
stemming  from the charge reconstruction of \pmt{} waveforms where several \pe{}s pile up.
To provide the community with the possibility to test our reconstruction algorithm on different inputs, and to compare
its performance to different reconstruction tools, we made a C++ implementation available at~\cite{website}.


\acknowledgments
We are extremely grateful to S. Jetter for useful comments and suggestions. 
This work was partially supported by Ministero dell'Istruzione, dell'Universit\`{a} e della Ricerca (MIUR) under the scientific project PRIN 278 2012 (CPPYP7-003), by National Institute for Nuclear Physics (INFN) through the JUNO experiment and ITALian RADioactivity project (ITALRAD), by Progetto Agroalimentare Idrointelligente Aladin CUP D92I16000030009, and by Universit\`{a} degli Studi di Ferrara under the scientific project FIR-2017.


\begin{thebibliography}{99}  

\bibitem{website}
C++ implementation of the charge reconstruction algorithm described in this manuscript available at \url{http://www.fe.infn.it/CRA}

\bibitem{borexino}
Borexino Collaboration, G. Alimonti et al., 
\emph{Neutrinos from the primary proton-proton fusion process in the Sun},
\emph{Nature} {\bf 512} (2014) 383-386. \\
\url{http://dx.doi.org/10.1038/nature13702} .



\bibitem{dayabay} 
Daya Bay Collaboration, F. P. An et al., 
\emph{Observation of Electron-Antineutrino Disappearance at Daya Bay},
 108, 171803
\emph{Phys. Rev. Lett.} {\bf 108}  (2012) 171803. \\
\url{http://dx.doi.org/10.1103/PhysRevLett.108.171803} .



\bibitem{doublechooz}
Double Chooz Collaboration, Y. Abe et al.,
\emph{Improved measurements of the neutrino mixing angle $\theta_{13}$  with the Double Chooz detector},
\emph{JHEP} {\bf 10} (2014)  086  [\emph{Erratum ibid} {\bf 02} (2015) 074 ]. \\
\url{http://dx.doi.org/10.1007/JHEP10(2014)086} .


\bibitem{kamland}
KamLAND Collaboration, A. Gando et al.,
\emph{Reactor on-off antineutrino measurement with KamLAND},
\emph{Phys. Rev. D} {\bf 88}  (2013) 033001.\\
 \url{https://doi.org/10.1103/PhysRevD.88.033001} .


\bibitem{reno}
RENO Collaboration, J.H. Choi et al., 
\emph{Observation of Energy and Baseline Dependent Reactor Antineutrino Disappearance in the RENO Experiment},
\emph{Phys. Rev. Lett.} {\bf 116} (2016)  211801. \\
\url{https://doi.org/10.1103/PhysRevLett.116.211801} .


\bibitem{juno}
JUNO Collaboration, Fengpeng An et al., 
\emph{Neutrino Physics with JUNO},
\emph{J. Phys. G: Nucl. Part. Phys.} {\bf 43}  (2016) 030401. \\
\url{http://dx.doi.org/10.1088/0954-3899/43/3/030401} .


\bibitem{jinping}
Wu, Yu-Cheng et al., 
\emph{Measurement of cosmic ray flux in the China JinPing underground laboratory},
 \emph{CPC} {\bf 37}  (2013) 086001. \\
\url{https://doi.org/10.1088/1674-1137/37/8/086001} .


\bibitem{reno50}
Soo-Bong Kim,
\emph{New results from RENO and prospects with RENO-50},
\emph{Nucl. Part. Phys. Proc.}  {\bf 265-266}  (2015)  93-98 . \\
\url{http://www.sciencedirect.com/science/article/pii/S2405601415003661?via\%3Dihub} .
 

\bibitem{snoplus}
SNO+ Collaboration, S. Andringa et al.,
\emph{Current Status and Future Prospects of the SNO+ Experiment},
\emph{High Energy Phys.} {\bf 2016}  (2016) 6194250.\\
\url{http://dx.doi.org/10.1155/2016/6194250} . 
 

\bibitem{andes}
P. A. N. Machado, T. Muhlbeier, H. Nunokawa, and R. Zukanovich Funchal,
\emph{Potential of a neutrino detector in the ANDES underground laboratory for geophysics and astrophysics of neutrinos},
\emph{Phys. Rev. D} {\bf 86} (2012) 125001. \\
\url{https://doi.org/10.1103/PhysRevD.86.125001} .


\bibitem{bowden}
N.S. Bowden, K.M. Heeger, P. Huber, C. Mariani, R.B. Vogelaar,  
\emph{Applied Antineutrino Physics 2015 - Conference Summary},
arXiv:1602.04759 (2016).


\bibitem{mcppmt}
Y. Wang, 
\emph{Large Area MCP-PMT and its Application at JUNO}, 
Talk at XVII  International Workshop on Neutrinos Telescopes (NeuTel 2017) , Venice, Italy, 2017, available at 
\url{https://agenda.infn.it/contributionDisplay.py?sessionId=15&contribId=55&confId=11857} .

\bibitem{dayabay_fadc}
Y. Huang et al,  
\emph{The Flash ADC system and PMT waveform reconstruction for the Daya Bay Experiment},
arXiv:1707.03699 (2017). 

\bibitem{jetter}
S. Jetter et al, 
\emph{PMT waveform modeling at the Daya Bay experiment},
\emph{CPC} {\bf 36}   (2012) 93-98.\\
\url{https://doi.org/10.1088/1674-1137/36/8/009} 

\bibitem{dossi}
R. Dossi, A. Ianni, G. Ranucci, O.Ju. Smirnov,
\emph{Methods for precise photoelectron counting with photomultipliers},
\emph{Nucl. Instrum. Methods Phys. Res. A } {\bf 451}  (2000) 623-637\\
\url{http://dx.doi.org/10.1016/S0168-9002(00)00337-5} .


\bibitem{chess}
J. Caravaca et al,
\emph{Experiment to demonstrate separation of Cherenkov and scintillation signals},
\emph{Phys. Rev. C} {\bf 95} (2017) 055801
\url{https://doi.org/10.1103/PhysRevC.95.055801}


\bibitem{steven}
Steven W. Smith , 
\emph{The Scientist and Engineer's Guide to Digital Signal Processing}, 
California Technical Publishing, San Diego, U.S.A. (1998). 

\bibitem{wiener_description_6}
S.V. Vaseghi,
\emph{Advanced Digital Signal Processing and Noise Reduction}, 
John Wiley \& Sons Ltd., Chichester U.K. (2000)

\bibitem{wiener_mit}
A.V Oppenheim, G.C. Verghese, 
\emph{Signals, Systems, and Inference}, 
\emph{Class notes ``Introduction to Communication, Control and Signal Processing''}, 
MIT 2010
\url{https://ocw.mit.edu/courses/electrical-engineering-and-computer-science/6-011-introduction-to-communication-control-and-signal-processing-spring-2010/readings/MIT6_011S10_chap11.pdf}


\end{thebibliography}
\end{document}